# The complementary contributions of academia and industry to AI research

Lizhen Liang[1], Han Zhuang[2], James Zou[3], Daniel E. Acuna[4,*]

[1]Syracuse University
[2]Ningbo Institute of Digital Twin
[3]Stanford University
[4]University of Colorado at Boulder

*Corresponding author: `daniel.acuna@colorado.edu`

**Abstract.** Artificial intelligence (AI) has seen fast paced development in industry and academia. However, striking recent advances by industry have stunned the field, inviting a fresh perspective on the role of academic research on this progress. Here, we characterize the impact and type of AI produced by both environments over the last 25 years and establish several patterns. We find that articles published by teams consisting exclusively of industry researchers tend to get greater attention, with a higher chance of being highly cited and citation-disruptive, and several times more likely to produce state-of-the-art models. In contrast, we find that exclusively academic teams publish the bulk of AI research and tend to produce higher novelty work, with single papers having several times higher likelihood of being unconventional and atypical. The respective impact-novelty advantages of industry and academia are robust to controls for subfield, team size, seniority, and prestige. We find that academic-industry collaborations produce the most impactful work overall but do not have the novelty level of academic teams. Together, our findings identify the unique and nearly irreplaceable contributions that both academia and industry make toward the progress of AI.

## INTRODUCTION

Artificial Intelligence (AI) has seen a dramatic surge in development, applicability, and perceived value over the last ten years (*1*, *2*). There are promising applications in almost all aspects of society, from healthcare to education to finance (*3*, *4*). Historically, most ideas originated in academia, but recent striking results in generative AI (e.g., ChatGPT and Midjourney) seem to overshadow academic AI research in favor of industry progress (e.g., see (*5*–*7*)). But how accurate are these perceptions? Are the advances from industry AI research so dominant that the research agenda of academic teams are challenged? Or are advances from industry getting the attention while academia publishes underappreciated novel work? And finally, are there academic–industry collaborations that would bring the "best of both worlds"? These questions motivate us to investigate how industry and academic teams differ in the publications and AI models they produce. Understanding the disparities of impact and innovation between the two environments could have important implications for further facilitating advancements in AI-related research fields and AI-related industries.

Academics were the leaders of early advancements in AI. For example, the groundbreaking implementation of the *Perceptron* was carried out by Frank Rosenblatt at the Cornell Aeronautical Laboratory (*8*); backpropagation was formally introduced by Rumelhart, Hinton, and Williams while working at UC, San Diego and CMU (*9*). More recently, the initial deep learning advances also came from academia. Hinton and Salakhutdinov (*10*) (University of Toronto) introduced efficient methods for training deep networks, and Fei-Fei Li (Stanford University) introduced the ImageNet dataset (*11*, *12*). Academics seem to have the freedom and flexibility to explore problems that the private sector does not yet value.

In recent years, however, industry research teams seem to have produced more breakthroughs. For example, Microsoft Research's ResNet (*13*), Google AI's Transformers and BERT (*14*, *15*), and OpenAI's GPT (*16–18*) are all examples of groundbreaking work that mostly originated in industry. There are obvious advantages that industry teams hold over academics. Deep learning models rely heavily on computational power and large datasets, which are readily available to industry (*19*). Due to this disadvantage, academic researchers sometimes prefer to focus on different problems instead (*7*, *20*). Additionally, how AI research is conducted and motivated varies significantly between industry and academia. In the private sector, scientific discoveries have historically been transformed into marketable products (*21*); academic discoveries usually do not have such clear mandates (*22*).

The availability of scientific research metadata offers an opportunity to study the differences in the research produced by academia and industry. In the present study, we investigate the impact and novelty of hundreds of thousands of articles from a dataset of AI conferences between 1995 to 2020. We also analyze hundreds of models from records of state-of-the-art results for natural language processing and computer vision tasks. We categorize co-authoring teams into industry, academic, and academic–industry teams. We show that industry teams tend to be better cited and more citation-disruptive than academic teams, and produce more state-of-the-art models. On the other hand, academic teams produce significantly more novel work (highly atypical and less conventional) than industry teams. After controlling for field, team size, seniority, and team prestige, these differences still remain significant. Surprisingly, the novelty disadvantage of industry does not seem to improve when collaborating with academics while the impact disadvantage of academic teams does improve by collaborating with industry researchers. In our conclusion, we find that industry and academic research contributions are different and challenging to replace: to advance AI forward, we need the nurturing and growth of both environments.

**RELATED WORK**

Industry increasingly influences the progress of AI research. In (*23*), the authors show this effect by examining a number of factors, including Ph.D. and research faculty hires, the computing needs of academia vs. industry models, and the growing number of publications by industry.

Industry companies produce large data during their business operations, hiring increasingly more AI talents, and having better computing infrastructure, all key for successful progress. In a more specific study of NLP (*24*), the authors quantified and characterized industry presence in that community, showing that the impact of industry is significant and fast-growing. Similar to (*23*), the authors expressed concerns that industry's dominance could lead to monopolies, since data, trained models, and experiment results used by industry research teams are sometimes not shared. The authors thus called for transparency in NLP research to prevent monopolization. In another study of AI from 1990 to 2014, the authors found that academic–industry collaboration in AI is surprisingly rare compared to other CS fields (*25*). In (*26*), the authors studied the opportunities of future AI research, and they suggest that industry researchers are more likely to push new AI techniques because investments are now increasingly accumulating in this area: there is still an order of magnitude difference between government and private investments in the area (*27*). Overall, there is limited investigation on the effects of academic vs. industry work on AI research projects, and the results are inconclusive. To the best of our knowledge, there are no studies analyzing the novelty nature of research in academic vs. industry.

Many other research studies have compared academic and industry research work in fields other than AI. Most of these studies tend to be country-specific. In the context of Italian science (*28*), the authors examined the citation impact of publications indexed in the Web of Science from 2010 to 2017. They found that most publications in the private sector are from collaboration with academic researchers. Moreover, publications from teams with only industry researchers had a lower impact than those with only academic researchers. However, publications resulting from collaborations between industry and academia had a greater impact. Other authors have shown other trends that seem inconclusive. For example, no significant differences in citations between private–public collaborative publications and solely public (academic) collaborations (*29*), significant differences but for international collaborations, significant differences overall (*30*, *31*). Thus, the academic–industry collaboration patterns can be disparate, requiring a specific analysis of AI.

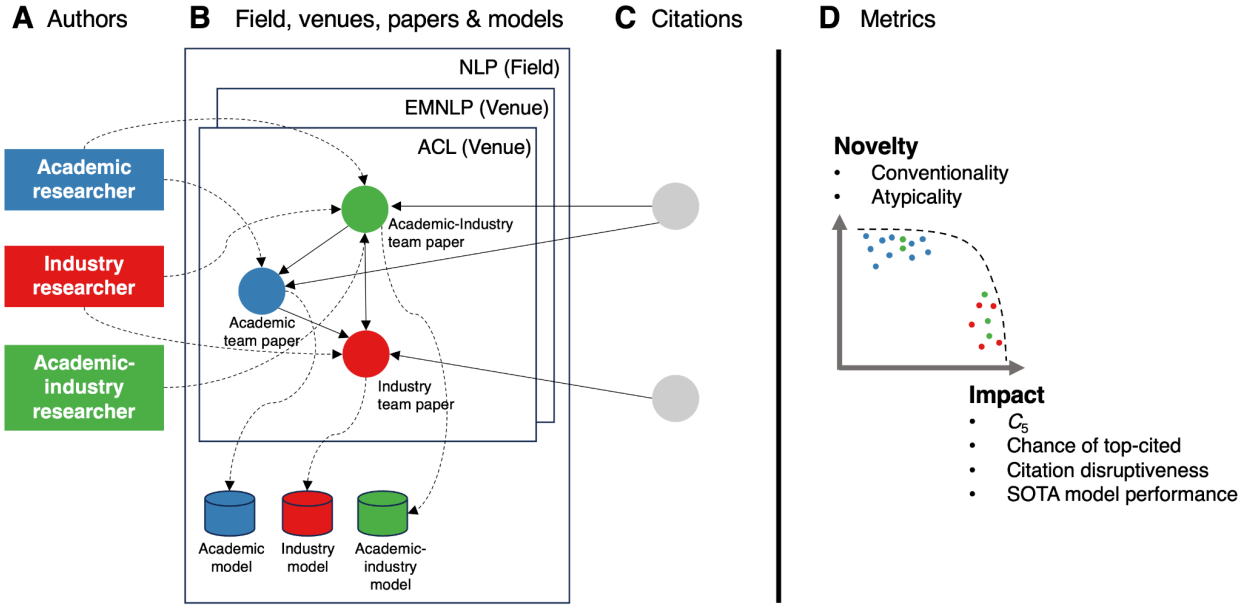

**Fig. 1. Academic articles and AI models emerge from collaborative efforts that are exclusively academic, exclusively industry, or a fusion of both, each presenting a distinct blend of impact and novelty. A.** Authors are classified as being academics or industry researchers. **B.** Papers are published in a venue (conference) and that venue belongs to a field. The papers published in such venues have teams that are exclusively academic (blue circles), exclusively industry (red circles), and academic-industry teams (green circles). Some of these papers published models for specific tasks (e.g., text summarization). Papers cite each other (continuous line with arrow) **C.** Papers can be cited by papers in the future (gray circles). **D.** Several metrics of a paper or a model produced by a paper are computing along the impact and novelty dimensions. Such metrics are analyzed across time, field, and team type.In general, we find that no team composition dominates both novelty and impact simultaneously.

## RESULTS

Our work analyzes the differences in novelty and impact between academic, industry, and academic-industry teams. We analyzed the affiliations of the researchers (Fig. 1A) publishing in conferences—the most common place to publish AI research. We grouped conferences into fields (Fig. 1B) and analyzed how the papers published by academic, industry, and academic-industry teams cite each other and how other publications cite them (Fig. 1C). We also analyzed the state-of-the-art models published by these types of teams. Finally, we analyze several metrics related to novelty and impact (Fig 1D). One of the ideas is to understand whether there are academic-industry arrangements that would dominate either industry or academia alone in the impact–novelty spectrum (Fig. 1D)

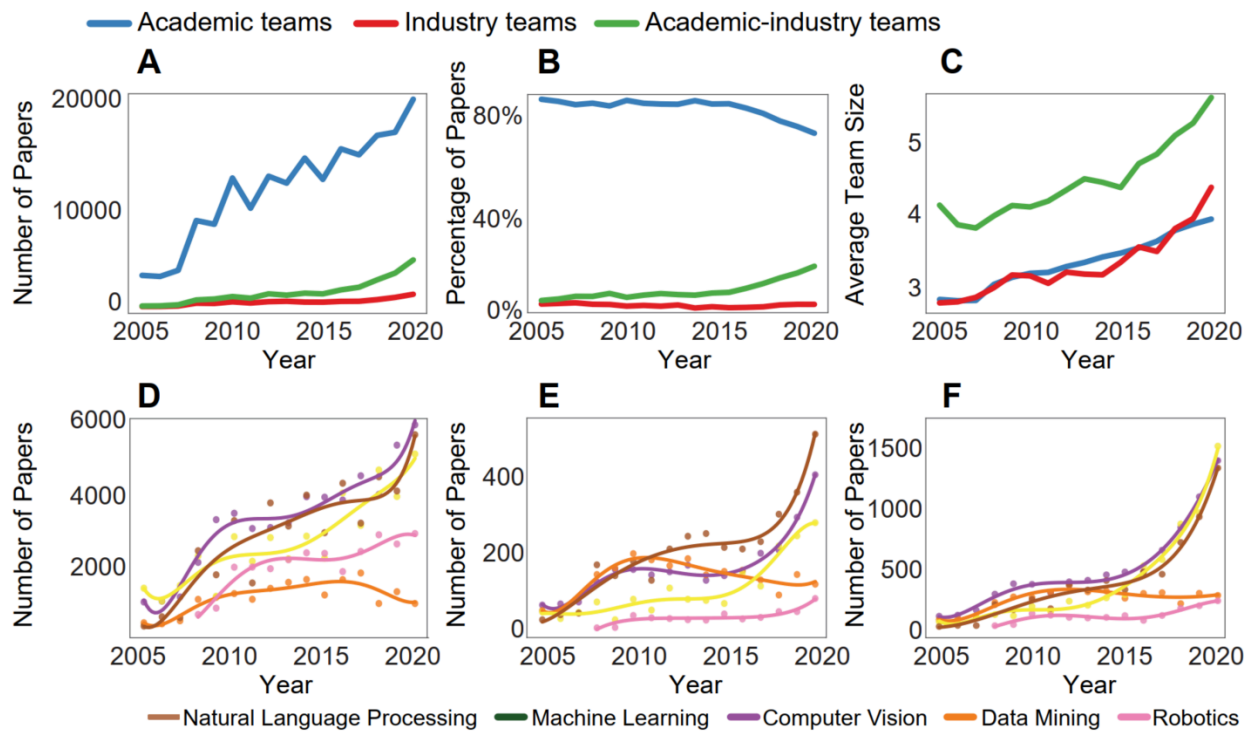

**Fig. 2. Publications count and team size are growing across team types and fields, with industry gaining momentum**; **A.** The number of papers published in AI publishing venues has been increasing over the years between 2005 and 2020 across different types of co-authorship teams; **B.** The percentage of publications from different types of co-authorship teams each year from 2005 to 2020; **C.** The size of co-authorship teams has been increasing over the years between 2005 and 2020 across different types of co-authorship teams; **D.** The number of papers published in different types of publishing venues over the years by teams with exclusively academic authors, **E.** The number of papers published in different types of publishing venues over the years by teams with exclusively industry authors; **F.** The number of papers published in different types of publishing venues over the years by teams with both industry and academic authors. The same figure but with years from 1995 to 2020 can be found in the Supplementary Materials (Fig S1), showing the same trend.

**Big science AI**

We start by analyzing the size of research in AI. Similar to general trends in science (*32–36*), AI has been growing rapidly. After dividing publications into academic teams, industry teams, and academic-industry teams, the number of publications by different kinds of teams has been steadily increasing since 2005 (Fig. 2A). In particular, there has been a marked increase since 2008. In terms of percentage (Fig. 2B), academic teams produce most of the work, but the number of publications from other kinds of teams started to increase. When looking at the team size (Fig. 1C) there has been a steady growth in the mean team size across different kinds of co-authorship teams, which aligns with the idea that science has become a team sport (*37*). Different subfields show a similar growing trend across team types (Figs 2D, 2E, and 2F are academic teams, industry teams, and academic-industry teams, respectively). We observe that the number of papers in the field of computer vision, machine learning, and natural language processing has increased the most compared to the other sub-fields of robotics, and data mining, regardless of the type of co-authorship teams. As expected, the field of NLP has seen a dramatic recent expansion across the board (Figs 2D-F), but computer vision is still the largest subfield.

**Impact metrics: highly cited and citation disruptive articles**

While the number of publications in the field of AI has been increasing over the years, some publications have attracted more attention from the community, being cited more often and disrupting the citation network. To better understand how teams might contribute to the disparity of research impact, we compared the likelihood of publishing a high-impact paper across publication years for different types of teams. Similar to previous work measuring impact (*38*), we calculated the citation each paper received within the five years after publication as a measure of impact. We further ranked these citation counts and identified papers that are among the top 10 percent most cited AI papers each year. We find that academic-industry teams (Fig. 3A) are more likely to produce high-impact publications than industry teams and academic teams—when we later control for team size and field, the advantage of these collaborative teams goes away. We observe that in the year 2020, citation counts for AI publications published by research teams with exclusively industry authors in 2015 were 73.78 percent more likely to be highly cited compared to research teams with exclusively academic authors ($t(116,536) = 3.41$, $p < 0.001$).

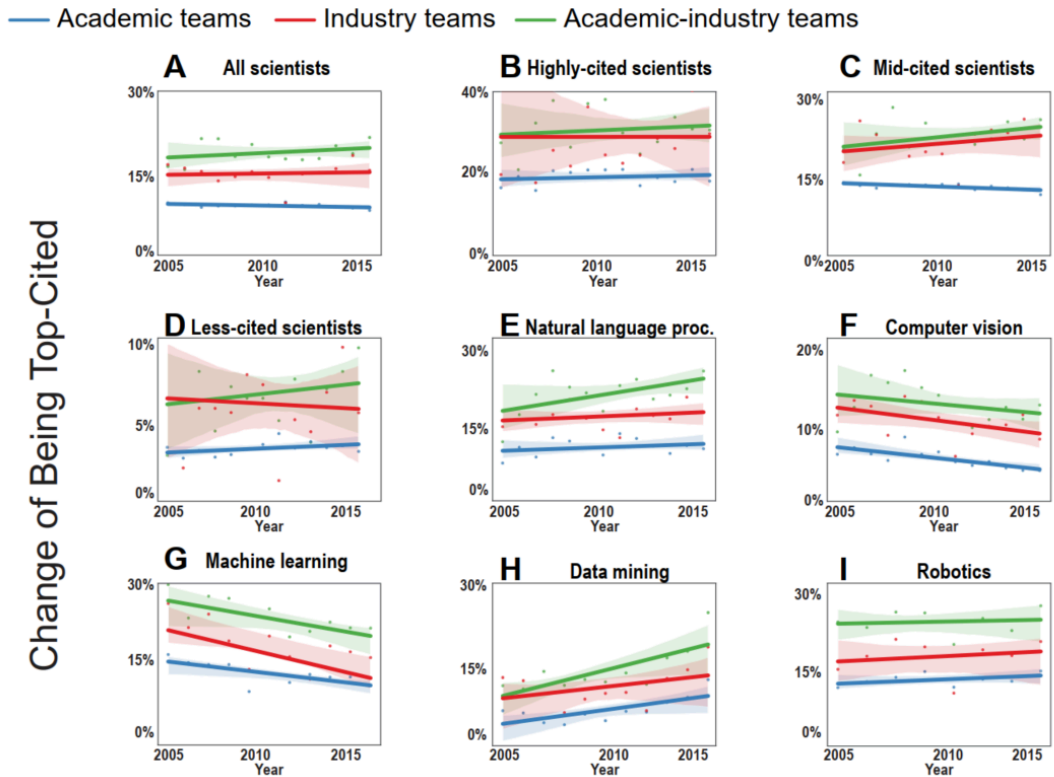

Fig. 3. Academic-industry teams outperform industry and academic teams in raw impact across all scientists, teams of similarly-cited authors, and fields, from 2005 to 2015; **A.** the chances of publishing top 10% cited AI papers between 2005 and 2015 for teams of different co-authorship types; **B.** the chances of publishing top 10% cited AI papers between 2005 and 2015 for teams of different co-authorship types for teams with top 10% team average h-index; **C.** the chances of publishing top 10% cited AI papers between 2005 and 2020 for teams of different co-authorship types for teams with top 50% team average h-index; **D.** the chances of publishing top 10% cited AI papers between 2005 and 2015 for teams of different co-authorship types for teams with bottom 25% team average h-index; **E.** the chances of publishing top 10% cited AI papers for papers in natural language processing conferences; **F.** the chances of publishing top 10% cited AI papers for papers in computer vision conferences; **G.** the chances of publishing top 10% cited AI papers for papers in machine learning conferences; **H.** the chances of publishing top 10% cited AI papers for papers in data mining conferences; **I.** the chances of publishing top 10% cited AI papers for papers in robotic conferences. The same figure but with years from 1995 to 2020 can be found in the Supplementary Materials (Fig S2), showing the same trend.

The high-impact of a publication could be the result of many factors, including the nature of authorship affiliation and whether authors themselves are influential in their field. We break all co-authorship teams into teams with high average h-index (teams within the top 10% highest team average h-index, Fig 3B), medium average h-index (teams with the top 50% highest team average h-index, Fig. 3C), and relatively low average h-index (teams with the bottom 25% highest team average h-index, Fig 3D). We found a consistent trend across these subgroups, suggesting that controlling for the team type still makes an industry team dominant.

The impact of research publications can also be influenced by the research field or the venue where the paper was published. We break down the field of AI into sub-fields. We use the categorization of AI conferences provided by the AI deadline website[1] (see Materials and Methods), grouping conferences into natural language processing (Fig. 3E), computer vision (Fig. 3F), data mining (Fig. 3G), machine learning (Fig. 3H), and robotics (Fig. 3I). Across fields, we observe the same trend that academic-industry teams are more likely to produce highly cited papers, followed by industry teams.

Finally, we estimated the (citation) disruptive index of AI articles from academic teams, industry teams, and academic-industry teams (see Methods for more details). Overall, we see a dramatic increase in disruptiveness across team types after 2010 (Fig. 4E). We also observed that the average citation disruptive index for industry publications is 42.44 percent higher than academic publications ($t(21,852) = 2.75$, $p < 0.001$). These results suggest that articles from academic teams are more likely to consolidate citations, but articles from industry teams or academic-industry teams are more disruptive. This might be because industry teams usually propose new neural network architectures, of which predecessors are outdated, while industry or academic-industry teams publish to interdisciplinary subfields, which are more likely to be cited by articles from another field, which propose ideas based on other fields.

---

[1] https://aideadlin.es/

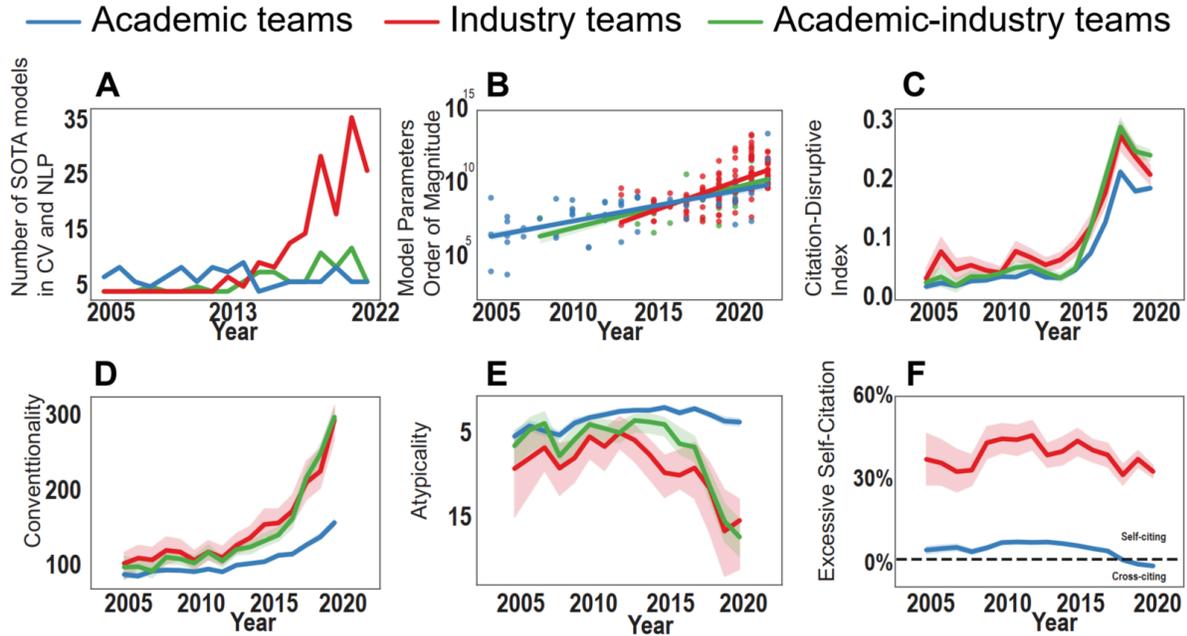

**Fig. 4. Industry and academic-industry teams lead in state-of-the-art models and citation disruptiveness while academia excels at novelty and cross-team citation behavior**; In **A** and **B**, models being visualized satisfy the following criteria: they have received at least 1000 citations; the research community highly accepts them; they received state-of-the-art performance at the time of publication; they have been deployed in large projects (*38*); In **F**, excess within-group citation is measured by negative outsize within-group citation difference in difference; **A.** the number of highly-cited state-of-the-time models published by teams of different co-authorship types; **B.** the number of model parameters (order of magnitudes) for teams of different co-authorship types; **C.** the citation-disruptiveness of publications published by teams of different co-authorship types; **D.** the conventionality of publications by teams of different co-authorship types; **E.** Atypicality of publications by teams of different co-authorship types (according to the original model by (*39*), lower atypicality is un-intuitively *more* novel; therefore, y axis is flipped to increase interpretability); **F.** the excess within-group citation between teams with only academic researchers and teams with only industry researchers. The same figure but with a longer time period can be found in the Supplementary Materials (Fig S3), showing the same trend.

**State-of-the-art status of AI models**

AI models with better performance are more likely to be adopted due to their state-of-the-art status, becoming more impactful in the research community. We selected 173 notable AI models in popular language and vision areas to analyze the gap between industry and academia. We use the models examined in (*39*), which satisfy the following criteria set by the authors: 1) they have received at least 1,000 citations, 2) the research community highly accepts them, 3) they received state-of-the-art performance at the time of publication, and 4) they have been deployed in large projects. The model dataset covers 7 decades, but almost 80% of them have been proposed during the last decade.

In our analysis, at the beginning of the deep learning era (circa 2012), most SOTA-performing AI models were still produced by academia (Fig. 4A). Most of these models had under 1 billion parameters in size (Fig. 4B). As an example, in 2013, Google published their Word2Vec (large) model with 6.92 billion parameters. After that, AI models proposed by the industry dominated both the field of natural language processing and computer vision in model size. We observed that in 2022, industry teams proposed 25 state-of-the-art models. In contrast, academic teams only proposed 2, showing that industry teams proposed 11.5 times more state-of-the-art models than academic teams (exact binomial test, $N = 27$, $K = 2$, proportion = 0.08, $p < 0.001$).

Models with better performance tend to have more parameters appears to be a common trend across different AI subfields. For instance, GPT-1 has 0.12 billion parameters, GPT-2 has 1.5 billion parameters, and GPT-3 has about 175 billion parameters (*16–18*). We noticed a significant growth in the size of AI models proposed by both industry and academic teams. Academics propose larger models compared with the other types of co-authorship teams before 2010, which is the pre-deep-learning era. However, after 2010, industry teams and academic-industry teams produced larger models. At the same time, the size of AI models increased immensely within a decade, as shown in Figure 4B.

**Novelty and conventionality over time in academia and industry AI research**

A second fundamental function of science is to advance innovation through the publication of novel research (*22*). Which kind of team produces the most novel research? To investigate this question, we estimated the *atypicality* and *conventionality* of AI articles from academic teams, industry teams, and academic-industry teams (see Methods for more details). These have been proposed as proxies for novelty. Briefly, atypicality measures how unusual are the most usual citations of an article; conventionality measures how unusual are the most common citations of an article. Intuitively, if atypicality and conventionality are high, it indicates novel combinations of knowledge.

Our results show that after 2010, AI articles from academic teams were more novel and less conventional (Fig 4D and Fig 4E, respectively) compared to AI articles from industry teams or academic-industry teams. Because atypicality is a highly skewed metric, if we look at the median of each type of team, we find that in 2020 the median atypicality of academic teams was 2.8 times higher than the median atypicality of industry teams ($t(11,959) = 6.69$, $p < 0.001$). These results suggest that academic teams are likely to pursue novel research., Although previous literature (*40*) shows that novel research is more likely to receive more citations than conventional research, the discipline of AI might have a different tradition, i.e., state of the art AI models can attract more citations than other types of AI research.

**Within and cross group citations across academic and industry AI research**

To understand how research proposed by different types of teams in AI would impact other types of teams, we investigate how different types of teams cite their own type of team or cross-cite other types of team. Figure 4F shows excess within-group citations between academic teams and industry teams. It shows that industry teams are more likely to cite papers by other industry teams, creating an "industry bubble." On the other hand, academic teams have less excess within-group citation, meaning that they are more likely to be familiar with literature from both groups.

**Field, team composition, and time controls for effect on impact and novelty**

This article investigated the relationship between paper impact, disruptiveness, novelty, and co-authorship types. Our previous analysis shows that academic-industry teams are more likely to produce high-impact papers, and academic teams are more likely to contribute novel ideas. While the results are straight-forward, they might obscure factors that might affect the impact and novelty of a paper, such as paper sub-fields, author seniority, co-authorship team size, average academic age (the year since the author first published a paper at the time, they published a newer paper), and the publishing year of the paper. For example, it could be possible that researchers in industry are highly accomplished already but researchers in academia are constantly renewing and more recent, creating a correlation between team type and impact. In this section, we fit mixed effect models to investigate further the simultaneous effect of these confounders on the impact and novelty of work.

**Table 1. Statistics for mixed effect models for impact.** Mixed effect models for predicting the impact of paper (log-citation counts 3 years after publishing). Model 1: Team characteristics random effects. Model 2: Model 1 + team size (fixed effect). Model 3: Model 2 + (academic age, h-index). Model 3: Model 2 + publication year fixed effects. Full model: Model 3 + field random effect. Models have increasingly better AIC, inciting a good fit in spite of model DF.

| Control variables | Model 1 | Model 2 | Model 3 | Full model |
|---|---|---|---|---|
| | \multicolumn{4}{c}{Log Citation Count within 3 years of publication ($C_3$)} | | | |
| *Fixed effects* | | | | |
| team size | | 0.0718*** | 0.08240*** | 0.07880*** |
| academic age | | | -0.00003 | -0.00001 |
| h-index | | | 0.03340*** | 0.02493*** |
| *Random effects* | | | | |
| Academic-industry teams | 0.263 | 0.207 | 0.163 | 0.156 |
| Academic teams | -0.225 | -0.202 | -0.178 | -0.170 |
| Industry teams | 0.038 | -0.005 | 0.015 | 0.014 |
| computer vision | | | | 0.167 |
| data mining | | | | 0.052 |
| machine learning | | | | -0.052 |
| natural language processing | | | | -0.076 |
| robotics | | | | -0.090 |
| Akaike's Information Criterion | 483900 | 482565.8 | 478088 | 476958 |
| Observations: | \multicolumn{4}{c}{149303} | | | |
| | \multicolumn{4}{c}{*: $p < 0.05$, **: $p < 0.01$, ***: $p < 0.001$} | | | |

*Mixed effect model of impact*

In this model, we set co-authorship team size, co-authorship team average academic age and h-index, and the year of publishing as independent variables. We then set research sub-field and co-authorship team types as mixed effects. We calculated the log-citation count three years after the publication of each paper as the dependent variable to measure impact. Aside from the full model considering all variables, we fitted three nested models considering a subset of variables to understand how each variable affects the impact of papers. In model 1, we set the co-authorship team type as a mixed effect, showing that without considering other factors, academic-industry teams are most likely to produce high-impact papers, followed by industry teams. Model 2 was created by adding co-authorship team size as an independent variable. In model 3, we added variables: average-academic age and h-index of the co-authorship teams for each paper to represent co-authorship team seniority. In the full model, we added publishing venues representing sub-fields as a mixed effect. Across model 2, model 3, and the full model, we found that academic-industry teams are more impactful compared to industry and academic teams, even though the differences in impact between academic-industry teams and industry teams becomes smaller as we consider different features of the co-authorship teams. The result from the full model also shows that papers in computer vision are more impactful, followed by data mining papers. We built the same model with logged citation counts five years after the publication of each paper and observed the same trend (see Table S1 in the Supplementary Materials).

**Table 2. Statistics for mixed effect models for novelty.** This models atypicality for different team, field, and paper characteristics. See Table 1 for model details. (+: in the original work by (*39*), higher atypicality meant less novelty, which strikes as non-intuitive. We use the *negative* atypicality to capture the intuition that higher values mean *more* novelty)

| Control variables | Model 1 | Model 2 | Model 3 | Full model |
|---|---|---|---|---|
| *Fixed effects* | | **Atypicality** | | |
| team size |  | 3.0620*** | 3.3360** | 2.9490*** |
| academic age |  |  | -0.0065 | -0.0008 |
| h-index |  |  | 0.638*** | 0.6817*** |
| *Random effects* | | | | |
| Academic-industry teams | 5.123 | 2.368 | 1.219 | 1.965 |
| Academic teams | 10.493 | 11.806 | 12.403 | 9.496 |
| Industry teams | -15.617 | -14.174 | -13.622 | -11.461 |
| computer vision |  |  |  | -6.271 |
| data mining |  |  |  | -4.877 |
| machine learning |  |  |  | 9.212 |
| natural language processing |  |  |  | -22.101 |
| robotics |  |  |  | 24.037 |
| Akaike's An Information Criterion | 2863035 | 2862946 | 2862870 | 2861821 |
| Observations | | 204786 | | |
| | | *P < 0.05, **p < 0.01, ***p < 0.0001 | | |

*Mixed effect model of novelty*

In this model, we set co-authorship team size, co-authorship team average academic age and h-index, and the year of publishing, as independent variables. We then set different types of publishing venues representing different research sub-fields in AI and co-authorship team types as mixed effects. We use the negative atypicality for each paper as the dependent variable to represent novelty. Like the mixed effect model for impact, we fitted three models using different subsets of variables aside from fitting a model with all the variables. Model 1 was fitted using only the co-authorship team types as mixed effects. The intercept for each group shows that without considering co-authorship team size, co-author seniority, and publishing venue, academic-industry teams are the most novel, followed by academic teams. After considering the effect of co-authorship team size, in model 2, the intercepts show that academic teams are the most novel, surpassing academic-industry teams, which is consistent with the result from the mixed effect models for impact that shows academic-industry teams are larger in size. After considering all factors, in the full model, the intercepts show that academic teams have produced research work with the most novelty compared with other types of co-authorship teams.

As shown in both table 1 and table 2, we see that team sizes matter for publication impact and novelty: publications with more co-authors possess more outlet of influence since each co-author will be the advocate for the paper. Teams with more co-authors are more likely to include researchers with different backgrounds, proposing more atypical ideas with such collaboration. In our result, we showed that collaboration between industry researchers and academic researchers are among teams with bigger publishing team size. Our statistical analysis also shows that different fields have different citing cultures, with some communities citing other work more often and other communities citing other work less often, causing the disparity in citation count between different subfields in AI.

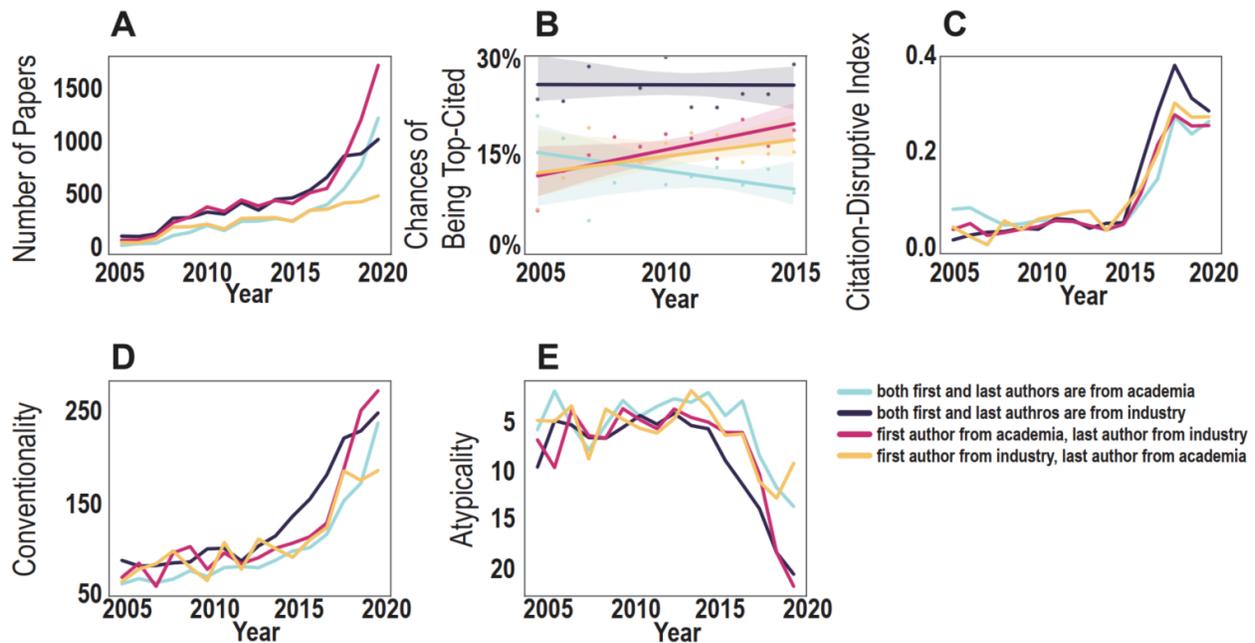

**Fig. 5. Academic-industry teams produce work they is similar to exclusively industry papers in impact and novelty; A.** the number of papers published by different kinds of academic-industry collaboration between 2005 and 2020; **B.** comparison of the impact of publications by different kinds of academic-industry collaboration between 2005 and 2020; **C.** citation-disruptiveness of papers published by different kinds of academic-industry collaboration between 2005 and 2020; **D.** atypicality of papers published by different kinds of academic-industry collaboration between 2005 and 2020; **E.** conventionality of papers published by different kinds of academic-industry collaboration between 2005 and 2020; **F.** the distribution of author's academic age at the time of publishing a paper; **G.** disruptiveness of papers published by different kinds of academic-industry collaboration between 2005 and 2020; **H.** Means and standard error of the mean of Gaussian mixture models clustering researcher h-index and academic age for authors published in different types of co-authorship. The same figure but with years from 1995 to 2020 can be found in the Supplementary Materials (Fig S2), showing the same trend.

**Investigating different types of academic-industry collaboration**

The previous results show that academic-industry teams are more likely to publish top-cited papers across different types of co-authorship teams and in different sub-fields in AI. We also observed that those collaborations produce publications that are more disruptive, less novel, and more conventional. To understand what makes those collaborations stand out in the field of AI, for papers produced by academic-industry teams, we further break them down into different groups based on whether their first authors and last authors are in the industry or academia. For papers with first or last authors affiliated with both industry and academic entities, we set those authors as academic authors. While there are 1,277 such cases out of 29,131 cases in the dataset we showed in Fig. 5. We found that results excluding those 1,277 cases are consistent with what we show in Fig. 5.

Fig. 5A shows the number of papers being published each year from 2005 to 2020 by different types of academic-industry collaborations. It shows that while counts of papers published by all types of teams are increasing over the years, teams with first authors from academia and last authors from the industry and teams with both academic first authors and last authors are increasing faster. When academics are driving the projects in academic-industry teams, we see the same characteristics as in academic teams: they have lower impact but higher novelty. Similarly, when industry teams drive the projects in mixed teams, they have high impact and low novelty.

For papers produced by academic-industry teams, with the same categorization we mentioned above, we were able to break this class into four different classes based on first and last author affiliation. We examine the percentage of high-impact papers for each type of collaboration. Among the four types of collaborations, we assume that the first author and the last author of the paper are the leading authors for those publishing co-authorship teams. Figure 4b shows that out of all types of collaborations, teams with both first authors and last authors from the industry are more likely to produce high-impact papers. In contrast, teams with first and last authors from academia are less and less likely to produce high-impact papers. Papers with the first author from industry but the last author from academia and papers with the last author from academia but the last author from industry are cases where both industry and academia authors are leading the project, with relatively lower impact compared to papers with both first and last authors from the industry.

**Disruptiveness, novelty, and conventionality for different types of academic-industry collaborations**

We investigate disruptiveness, novelty, and conventionality indexes to get a better understanding of how those indexes change over time for different types of academic-industry teams, we calculated those indexes for those teams and plotted the averaged index by year. Fig. 5C shows the disruptiveness index of papers published by different types of academic-industry teams. Consistent with Fig. 4C, teams with both first and last authors from industry shared traits with teams that are exclusively industry and have the most disruptiveness among the four groups, while teams with both first and last authors from academia have relatively low disruptiveness. Fig. 5D and 5E show the conventionality and atypicality of different types of academic-industry collaborations. Fig. 5E shows that teams with first authors from academia and last authors from industry, along with teams with both first and last authors from industry, are more conventional. 5D shows that teams with the first author from industry and the last author from academia are more atypical, followed by teams with both the first and last authors from academia.

Our analysis shows that among different types of collaboration between industry and academia, teams led by industry researchers, with both first and last authors from the industry, have produced more impactful and disruptive publications, while teams led by academic researchers, with first and last authors from academia, are producing less impactful papers, indicating that teams led by academic researchers might not be benefitted by their collaboration with industry researchers. However, our analysis also shows that teams with both first and last authors from academia and teams with academic last authors and industry first authors have published papers that are less conventional and more atypical, which are similar to teams of exclusively academic authors. A qualitative analysis of the top highly cited papers, citation-disruptive, and novel papers (Supplementary material) shows that highly atypical work usually combines disparate fields (e.g., "Learn To Be Uncertain Leveraging Uncertain Labels In Chest X Rays With Bayesian Neural Networks" published in a CVPR workshop combines Bayesian statistics with neural works with an application to radiology), and highly impactful are articles (e.g., "Adam: A Method for Stochastic Optimization" published in ICLR introduced one of the most common optimizers used for training neural networks today).

**Co-authors for different types of academic-industry collaborations**

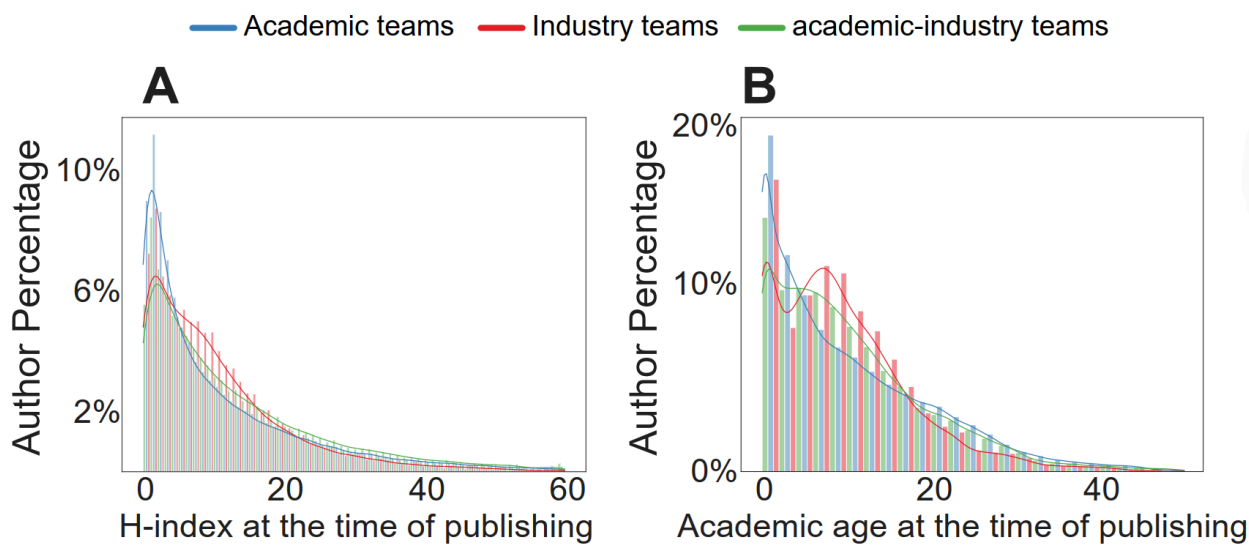

**Fig. 6.** Academic age and h-index of authors at the time of publishing with different types of co-authorship teams.

To have an overview of the researcher population, we plot the histogram of researchers' h-index and academic age at the time they published a paper after (including) the year 2000, shown in Fig. 6. For h-indexes, we show the distribution of h-indexes at the time each author published an AI paper, with h-indexes less than 100. We show that across different types of teams, the distribution of h-indexes generally follows the same trend. We also show the distribution of academic age at the time each researcher published an AI paper for researchers with less than 50 years of academic age. We found that for authors published in industry-academic collaboration teams, there is a peak around 4-6 years of academic age. While for authors published in all industry teams, there is a peak at 7-8 years of academic age. To further compare the seniority of authors publishing in different types of co-authorship teams, we train Gaussian mixture clustering models on different types of teams with h-index and academic age at the time of publishing. We set the number of components to two and compared the centroid of the less senior cluster and the more senior cluster across different types of co-authorship teams. For authors published in academic teams, the less senior cluster has a centroid of 3.22 h-index and 3.35 academic age, while 23.28 h-index and 19.00 academic age for the more senior cluster. For authors published in industry teams, the less senior cluster has a centroid of 5.40 h-index and 5.35 academic age, while 22.91 h-index and 17.60 academic age for the more senior cluster. For authors published in academic-industry teams, the less senior cluster has a centroid of 6.71 h-index and 5.36 academic age, while 31.65 h-index and 19.87 academic age for the more senior cluster. By comparing these centroids, we observe that the average seniority for both more senior and less senior co-authors in industry-academia collaboration teams is more senior compared with other types of teams.

**Table 3. Mean and SEM of H-index and academic age for junior and senior team members.** The h-index and academic age of junior researchers in industry are higher than in academia.

| Co-Authorship Team Types (Researcher Affiliation) | H-index (junior) | Academic age (junior) | H-index (senior) | Academic age (senior) |
|---|---|---|---|---|
| Academic teams (all types of researchers) | 3.22 ± 0.003 | 3.35 ± 0.0039 | 23.28 ± 0.0218 | 19 ± 0.0122 |
| Industry teams (all types of researchers) | 5.40 ± 0.023 | 5.35 ± 0.0229 | 22.91 ± 0.0796 | 17.60 ± 0.0468 |
| Academic-industry teams (all types of researchers) | 6.71 ± 0.014 | 5.36 ± 0.0117 | 31.65 ± 0.0582 | 19.87 ± 0.0272 |
| Academic-industry teams (academics researchers) | 4.79 ± 0.013 | 3.85 ± 0.0116 | 30.88 ± 0.0810 | 19.77 ± 0.0366 |
| Academic-industry teams (industry researchers) | 9.16 ± 0.028 | 7.13 ± 0.0206 | 32.36 ± 0.0814 | 19.69 ± 0.0410 |

**DISCUSSION**

Motivated by the striking new advances in AI research by industry over the last years, we wanted to investigate whether and how AI research differs when it is produced by academic vs. industry teams. In particular, we used publication, citation, and state-of-the-art models published by these types of teams to investigate the impact, citation disruptiveness, novelty, and state-of-the-art status of their work. Our study indicates that research publications from industry-affiliated teams are generally more impactful, which is consistent with the results shown in (*30*) and (*28*) (but see (*29*)). We also show that publications from exclusively academic teams are less conventional and more atypical, suggesting that they are more novel and likely to introduce new, exploratory ideas. Our analysis reveals that the unique strengths of academic and industry teams are difficult to replace by academic-industry partnerships; collaborations in fact behave similar to industry teams.

We speculate that industry research is becoming increasingly more impactful and citation-disruptive because they have more access to data, computational power, and people. In the age of deep learning and big models, having large training datasets and modern computing hardware is crucial for advancing AI. Many tech companies rely on data and computing to drive their business, giving them a competitive advantage (*19*). In recent years, more and more Ph.D. students and professors have been influenced by industry research due to their abundant computational power, large datasets, and higher pay (*23*). Thus, the result from industry AI work is more prominent.

Academic teams seem to be producing the most novel research. This is consistent with a view in which academia addresses problems that industry does not consider (*22*). However, we also found that academic teams produced less impactful research. This paradoxical high-novelty–low-impact combination is unusual in the literature. For example, in the original atypicality and conventionality metrics by (*40*), they found that novelty is predictive of impact. Similarly, in the work of (*41*), they found that highly novel papers are less cited early in their existence but then reach equal or higher citations than more conventional work. We found that this citation disadvantage by academic teams did not diminish over time (Supplementary materials, Fig. S1.D). We can entertain a number of hypotheses. One is that the AI field moves very fast and that publications that are not "discovered" in time miss out citations due to fast obsolescence rates (*42*). Another hypothesis is that the influence of these publications is not perceived through direct citations—our metric of choice—but rather through network effects, which may be captured by network-based impact metrics such as the Eigenfactor (*43*). In the future, we will investigate these other explanations for the apparent divergence between impact and novelty.

We discovered that academic–industry teams are more impactful compared with other kinds of co-authorship teams. However, these teams have markedly different sizes, which could contribute to the impact differential of their papers. When controlling for team size, academic age, prestige, year, and field, we found that this academic–industry advantage diminished. Additionally, we found that the most impactful academic-industry teams are those where the first and last authors are from industry while the most novel ones are those where the first and last authors are from academia. Because first and last authors carry the bulk of the research guidance, this indicates that academic and industry researchers carry over their research cultures. A similar pattern is observed in the novelty space. In sum, teams that consist of a mix of both do not benefit from the competitive advantages of each culture.

Our study has some limitations. We apply citation-based metrics such as citation-disruptiveness, atypicality, and conventionally to understand industry and academic research. Even though those metrics are widely used in the field of science of science (*44*), they might not truly capture the disruptiveness and novelty in the colloquial sense of the word—we only care about citations and not other forms of disruption or novelty such as real technological advances (*22*, *45*). Other limitations include our controls for the mixed effect models (see Results). For example, the international composition of a co-authorship team or whether the team is formed through cross-institutional collaboration can also influence impact and novelty. Measuring those factors is beyond the scope of this study and should be considered for future research. Finally, we are not accounting for the effect of mobility on scientists. After all, most industry research is done by scientists who first trained in academia. The mentorship and research culture they enjoyed while being in academia is something that is not captured by citation information (*46*).

Despite its limitations, our study investigates a critical missing piece from the discussions about AI in academia and industry. While others have suggested that students in academia should move away from problems studied in industry (*7*), our results suggest that both environments are pursuing important work and need each other. Therefore, we would suggest that students in academia continue their research and intensify their efforts on pursuing novel and interesting work. In some fields, for example, industry does not have a culture of publishing results because it opens them up for intellectual property issues (*47*, *48*). Our study thus contributes to the understanding of the relationship between industry and academia for research in non-AI fields as well.

## CONCLUSION

This article aimed to understand the impact and novelty characteristics of different academic and industry team compositions. We found that academic teams have less impact than academic-industry and industry teams but produce significantly more atypical and, therefore, novel work. By analyzing academic-industry collaboration teams, we found that while they produce the highest impact compared to purely industry or academic teams, they cannot seem to enjoy the benefits of pure teams alone.

In future work, we will investigate how the individuals who are part of the teams might have an outsized effect on the impact and novelty of the articles. In our work, we only analyzed individuals as part of teams, and all our measures of performance were team-related. Also, we plan to understand whether there are significant variations across nations. Given the giant push for AI across the US, China, and Europe, there might be significant differences due to the disparate incentives in academics, industry, and academic-industry teams across these populations.

## ACKNOWLEDGMENTS

We would like to thank Chao Zhou for his help with the state-of-the-art analyses. DA would like to thank NSF grant #1933803. DA and LL were partially funded by Alfred P. Sloan's grant "Does Government Funding Change What You Do? The Effects of Funding on the Direction and Impact of Academic Energy Research".

**MATERIALS AND METHODS**

**Data**

*MAG*

Microsoft Academic Graph (MAG) includes rich information about many entities in the scientific process, including research papers, publishing venues, authors, affiliations, and fields. Using a copy of MAG we retrieved in late 2021, we analyze the metadata, including reference, authorship, and author affiliations, for research publications from AI publishing venues. We were able to identify conferences mentioned in the website "AI Conference Deadlines"[2], which includes the most active and popular conferences in the fields of AI and the website also includes categories for conferences, such as natural language processing (NLP), computer vision (CV), and data mining (DM).

To identify researchers affiliated with industry entities and differentiate them from researchers from academia, we first match affiliations from MAG with Research Organization Registry[3] to match affiliations that are from the industry. On top of that, we compiled a list of words that indicate the academic status of an affiliation, such as "university," "academy," "school," "college," and "faculty." We were able to identify the status (whether it is from the industry or academia) of the affiliation of an author at the time the author participated in the authorship of an AI paper.

*AI Conference Deadlines*

In order to identify AI papers, we selected representative AI conferences according to a website called "AI Conference Deadlines"[4], which lists popular AI conferences and their deadlines for paper submission. The website provides categorization for different AI conferences, which we used for identifying different sub-fields in AI. The mapping between subfields and conferences are shown in (table 4).

*Dataset for State-of-the-art and Model Sizes*

To keep track of the trend of state-of-the-art models in AI, we selected 173 notable AI models from (*39*) to analyze the gap between industries and academia. In (*39*), the authors selected AI models if they satisfied at least one of the following: 1. They have at least 1000 citations. 2. The research community highly accepts them (papers that were cited by other work as seminal). 3. They received state-of-the-art performance at the time of publication. 4. They have been widely deployed in large projects and important contexts, such as popular search engines and translation software. It's a curated collection of the SOTA AI models (*39*), which has comprehensive features of models from the 1950s to 2022. The dataset consists of models belonging to two domains, vision and language, and is categorized into three organization types: academia, industry, and academic-industry collaboration.

---

[2] https://aideadlin.es/
[3] https://ror.org/
[4] https://aideadlin.es/

**Methods**

*Disruption*

While there are several popular definitions of originality, disruption is one definition of originality that means "making previous state-of-the-art obsolete." The disruptiveness of a research article could be measured by what the paper cites and what papers are citing this paper: If a paper is cited by many papers that are not also citing what the paper cites, it shows that this paper makes what it cites obsolete. Thus, (*49*) proposed a network-based metric, the CD index, to measure the disruptiveness of a research paper, ranging from -1 to 1. If a research paper has a CD index of 1, then this paper is highly disruptive, and all papers citing this paper don't cite their predecessors. If a research paper has a CD index of -1, then this paper is not disruptive but consolidating, and all papers citing this paper also cite its predecessors. In this work, we compute these metrics to estimate the disruptiveness of AI papers to better understand the differences between research work done in academia and research work done in industry.

Table 4. Mapping between AI conferences and AI sub-fields

| Sub-fields | Conferences |
|---:|---|
| natural language processing | ACL, EMNLP, EACL, COLING, NAACL, IJCNLP, CONLL, LREC |
| computer vision | MM, ICML, ICMR, ACCV, 3DV, ICME, ICIP, MIDL, MICCAI, ICCV, FG, WACV, CVPR, ECCV, BMVC |
| machine learning | ECAI, NeurIPS, UAI, IJCAI, ICML, AISTATS, ICLR, ACML, AAAI, ALT, COLT, ICCC, ICPR, AAMAS, ICAPS |
| data mining | SIGIR, ICDM, PAKDD, PKDD, WSDM, CIKM, SDM, ResSys, WWW, ICWSM, KDD, ISMIS |
| robotic | IROS, RSS, HRI, ICRA |

*Novelty and conventionality*

To better understand how research work from the industry might be different from the work from academia, it is important to also look at how novel and conventional their work is, and in general, whether work from the industry is more novel compared to the work from academia, or is it the opposite? In this work, we used the method proposed by (*40*)(*49*) to estimate the novelty and conventionality of research work from the field of AI. This method quantifies the novelty and conventionality of a research paper through the novelty of combinations of cited journal pairs. If a paper's top 10% novel combination of publishing venues is novel across all papers in the previous year, then this paper can be considered a novel. If a paper's top 50% novel combination of publishing venues is conventional across all papers in the previous year, then this paper can be considered conventional. For each venue combination, we compute the likelihood of its appearance compared to a random co-reference network using a z-score. We measure the conventionality of a research paper by using the median of z-scores of all co-reference pairs of a paper; we then measure the novelty of a paper by looking at the top 10 percentile of z-scores of all co-reference pairs. If a co-reference has a negative z-score, then this pair of co-reference is novel; otherwise, it is conventional.

*Mixed effect model of impact across fields*

While the types of different teams would affect how impactful a work would be, for instance, people might cite papers affiliated with large companies because they seem more reliable, different natures linked to a research paper might also affect how likely a paper will be impactful, such as the size of the co-authorship team, the targeting field of the research paper. To show how decisive team type might be linked to how likely a paper will be impactful and how team type as a factor can be compared with other factors such as paper field, team size, and co-authorship team seniority, we fitted a mixed effect linear regression. Similar to linear regression, mixed-effect linear regression shows the relationship between the dependent variable and the independent variables and the relationship between independent variables. Different from linear regression models, mixed effect linear regression models assume that the dataset for fitting the model could be broken down into different groups, with relationships between variables differing in different groups. In the mixed-effect linear regression model we fit for this work, the field of the papers and the types of co-authorship teams are set as fixed effects.

*Excess self-citation analysis*

To further understand how publications could shape the research field by contributing ideas to teams from different environments, we measure the outsize self-citation using a difference-in-difference metric. We define excess self-citation as

$$ECC_t(I) = (P(I \mid I, t \leq t_p) - P(I \mid t \leq t_p)) - (P(A \mid I, t_p \leq t) - P(A \mid t \leq t_p))$$

$$ECC_t(A) = (P(A \mid A, t \leq t_p) - P(A \mid t \leq t_p)) - (P(I \mid A, t \leq t_p) - P(I \mid t \leq t_p))$$

where $ECC_t(I)$ refers to the outsize self-citation difference in difference percentage at time *t* of papers by exclusively industry teams citing papers by exclusively academic teams, while $ECC_t(A)$ refers to outsize self-citation difference in difference percentage of papers by exclusively academic teams citing papers by exclusively industry teams. $t_p$ is the time of publication of the papers considered. $P(α | β, t ≤ t_p)$ refers to the probability of paper in team type β to cite a paper from team type α while $P(α | t ≤ t_p)$ refers to the probability of any paper by any team type to cite a paper by team type α. Intuitively, these expressions correct for differences in cross-group citation and within-group citation cultures across different team types.

# Supplementary Materials for

# "The complementary contributions of academia and industry to AI research"

Lizhen Liang, Han Zhuang, James Zou, Daniel E. Acuna

Corresponding author: daniel.acuna@colorado.edu

The PDF file includes:

- Materials and Methods
- Supplementary Text
- Figs. S1 to S5
- Tables S1

**Supplementary material**

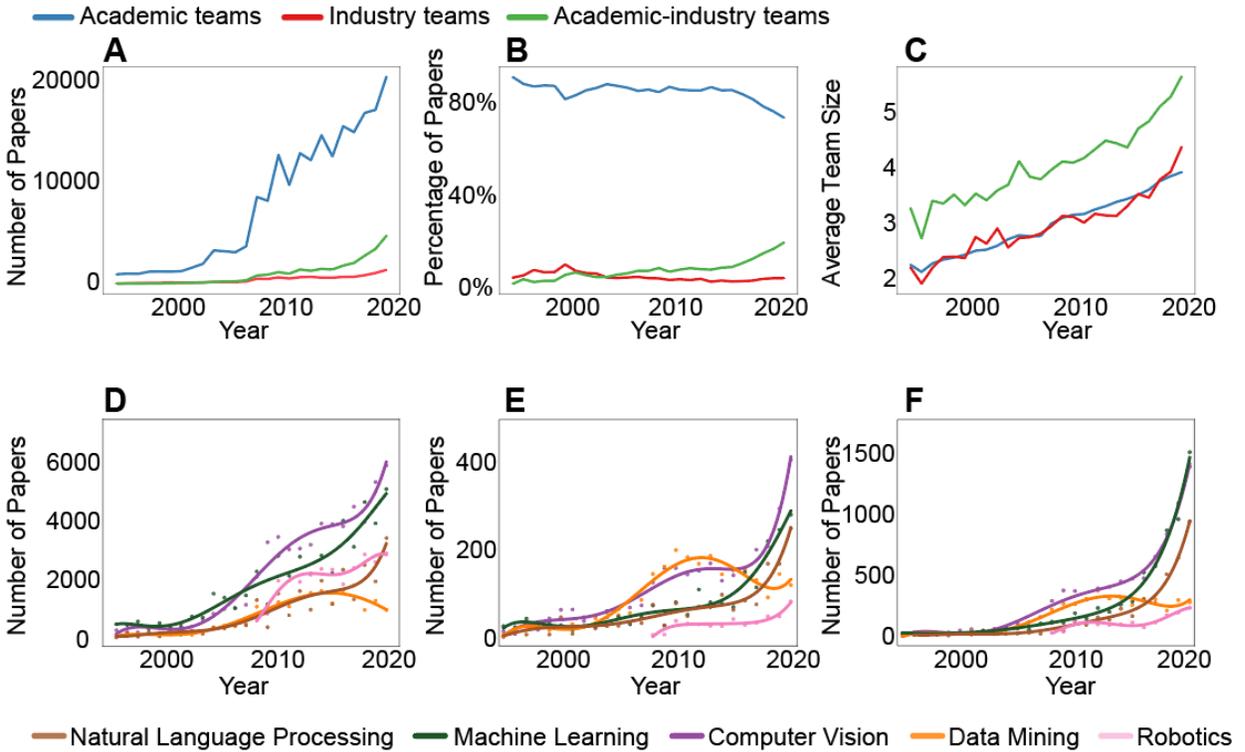

**Fig. S1. Publications count and team size are growing across team types and fields, with industry gaining momentum**; **A.** The number of papers published in AI publishing venues has been increasing over the years between 1995 and 2020 across different types of co-authorship teams; **B.** The percentage of publications from different types of co-authorship teams each year from 1995 to 2020; **C.** The size of co-authorship teams has been increasing over the years between 1995 and 2020 across different types of co-authorship teams; **D.** The number of papers published in different types of publishing venues over the years by teams with exclusively academic authors, **E.** The number of papers published in different types of publishing venues over the years by teams with exclusively industry authors; **F.** The number of papers published in different types of publishing venues over the years by teams with both industry and academic authors.

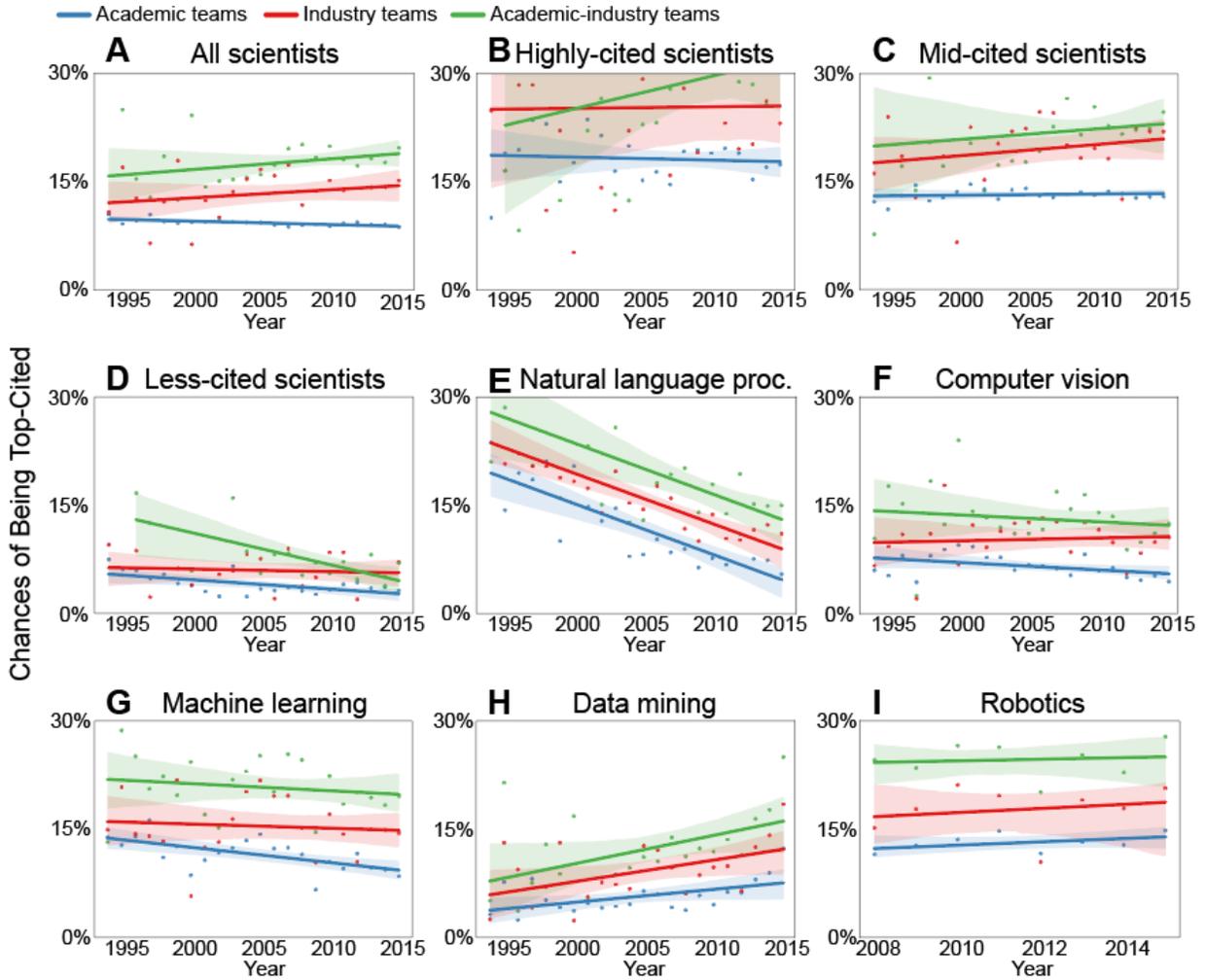

**Fig. S2. Academic-industry teams outperform industry and academic teams in raw impact across all scientists, teams of similarly-cited authors, and fields, from 1995 to 2015**; **A.** the chances of publishing top 10% cited AI papers between 1995 and 2015 for teams of different co-authorship types; **B.** the chances of publishing top 10% cited AI papers between 1995 and 2015 for teams of different co-authorship types for teams with top 10% team average h-index; **C.** the chances of publishing top 10% cited AI papers between 1995 and 2020 for teams of different co-authorship types for teams with top 50% team average h-index; **D.** the chances of publishing top 10% cited AI papers between 1995 and 2015 for teams of different co-authorship types for teams with bottom 25% team average h-index; **E.** the chances of publishing top 10% cited AI papers for papers in natural language processing conferences; **F.** the chances of publishing top 10% cited AI papers for papers in computer vision conferences; **G.** the chances of publishing top 10% cited AI papers for papers in machine learning conferences; **H.** the chances of publishing top 10% cited AI papers for papers in data mining conferences; **I.** the chances of publishing top 10% cited AI papers for papers in robotic conferences.

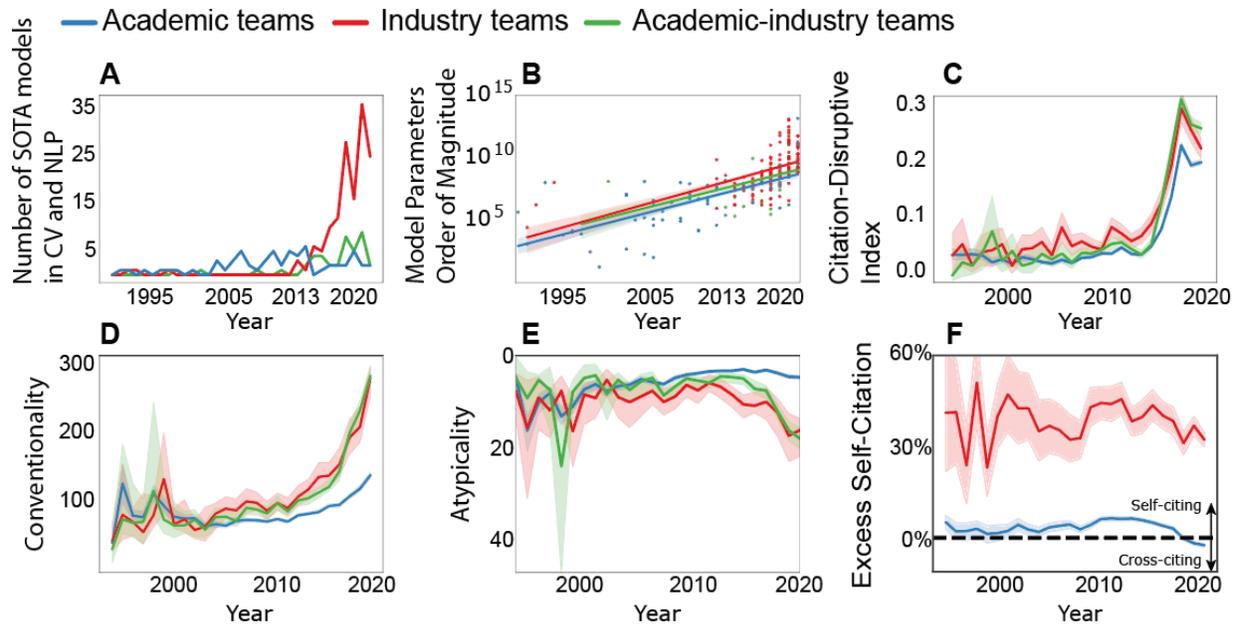

**Fig. S3. A.** the number of highly-cited state-of-the-time models published by teams of different co-authorship types; **B.** the number of model parameters (order of magnitudes) for teams of different co-authorship types; **C.** the citation-disruptiveness of publications published by teams of different co-authorship types; **D.** the conventionality of publications by teams of different co-authorship types; **E.** Atypicality of publications by teams of different co-authorship types, lower atypicality is un-intuitively *more* novel; therefore, y axis is flipped to increase interpretability); **F.** the excess within-group citation between teams with only academic researchers and teams with only industry researchers.

**Table S1. Statistics for mixed effect models for impact.** Mixed effect models for predicting the impact of paper (log-citation counts 5 years after publishing). Model 1: Team characteristics random effects. Model 2: Model 1 + team size (fixed effect). Model 3: Model 2 + (academic age, h-index). Model 3: Model 2 + publication year fixed effects. Full model: Model 3 + field random effect. Models have increasingly better AIC, inciting a good fit in spite of model DF.

| Control variables | Model 1 | Model 2 | Model 3 | Full model |
|---|---|---|---|---|
| | | Log Citation Count within 5 years of publication ($C_5$) | | |
| *Fixed effects* | | | | |
| team size | | 0.0594*** | 0.05121*** | 2.6820*** |
| academic age | | | -0.0008 | 0.1167 |
| h-index | | | 1.7610 | -0.0174*** |
| year of publishing | | | | 1.7390 |
| *Random effects* | | | | |
| Academic-industry teams | 0.248 | 0.201 | 0.130 | 0.123 |
| Academic teams | -0.210 | -0.192 | -0.169 | -0.149 |
| Industry teams | -0.038 | -0.009 | 0.039 | 0.027 |
| computer vision | | | | 0.080 |
| data mining | | | | 0.064 |
| machine learning | | | | -0.070 |
| natural language processing | | | | 0.085 |
| robotics | | | | -0.159 |
| Akaike's Information Criterion | 378997 | 378462.2 | 373470.4 | 372940 |
| Observations: | 112630 | | | |

*: $p < 0.05$, **: $p < 0.01$, ***: $p < 0.001$

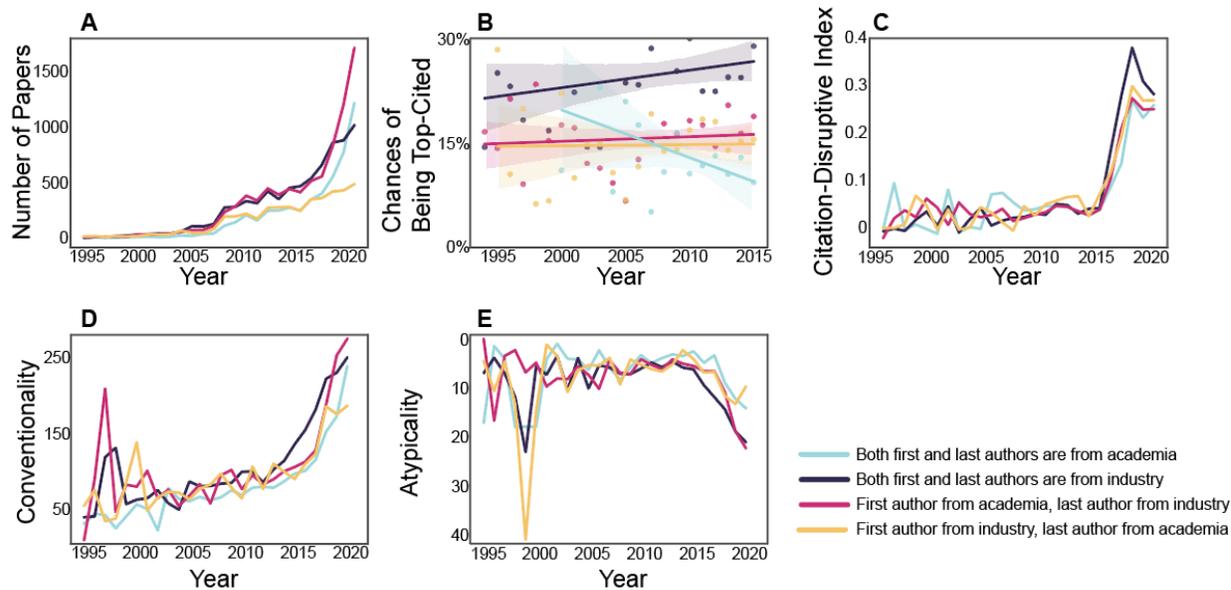

**Fig. S4. Academic-industry teams produce work they is similar to exclusively industry papers in impact and novelty; A.** the number of papers published by different kinds of academic-industry collaboration between 1995 and 2020; **B.** comparison of the impact of publications by different kinds of academic-industry collaboration between 1995 and 2020; **C.** citation-disruptiveness of papers published by different kinds of academic-industry collaboration between 1995 and 2020; **D.** atypicality of papers published by different kinds of academic-industry collaboration between 1995 and 2020; **E.** conventionality of papers published by different kinds of academic-industry collaboration between 1995 and 2020; **F.** the distribution of author's academic age at the time of publishing a paper; **G.** disruptiveness of papers published by different kinds of academic-industry collaboration between 1995 and 2020; **H.** Means and standard error of the mean of Gaussian mixture models clustering researcher h-index and academic age for authors published in different types of co-authorship.

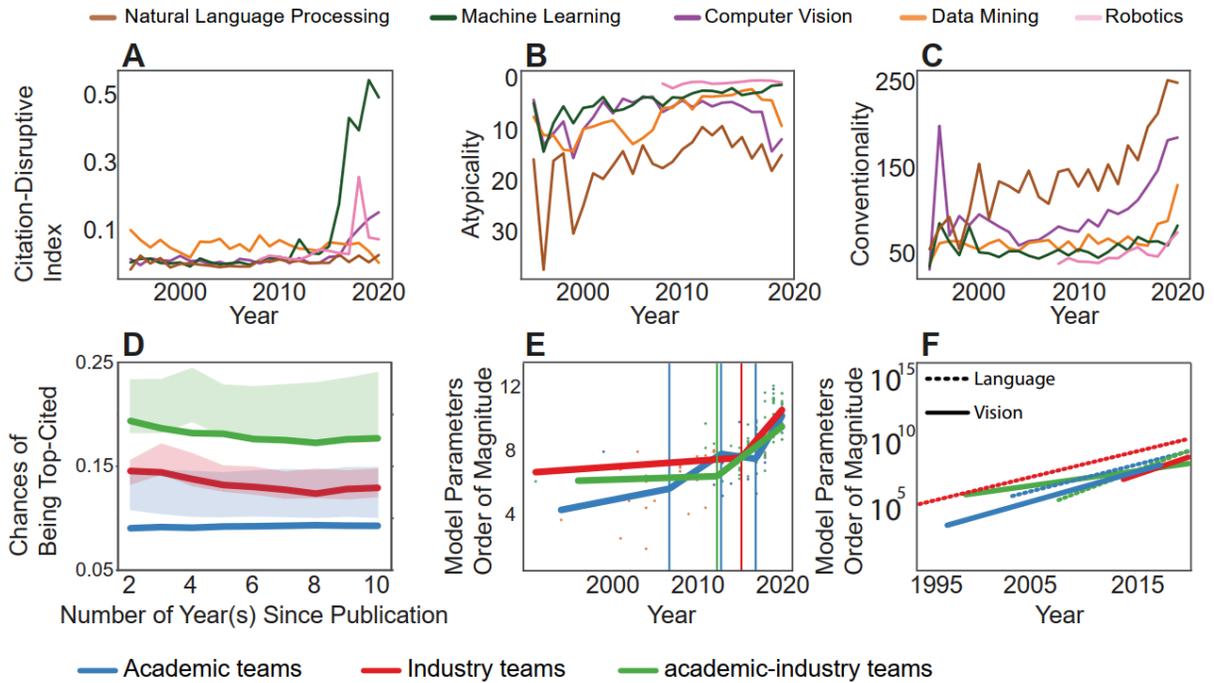

**Fig. S5. A.** Citation-disruptive index for papers published in different sub-fields represented by different publishing venues; **B.** Atypicality for papers published in different sub-fields represented by different publishing venues; **C.** Conventionality for papers published in different sub-fields represented by different publishing venues. **D.** the chance of publishing top 10% cited AI paper measured by citation count for a given number of year(s) since publication; **E.** piecewise regression on models parameter number. Both academia and industry models increased their parameters during 2013 - 2015. The rapid growth of industry started almost two years earlier than academia. **F.** the number of model parameters (order of magnitudes) for teams of different co-authorship types in the AI subfield of language and vision

Fig. S5. shows the citation disruptive index, atypicality, and conventionality across different sub-fields of different publishing venues. The mapping between sub-fields and publishing venues is shown in (Tabe 4.). In Fig. S5. B, we show that while robotics is the most atypical field, natural language processing is less atypical, and as shown in C, the field of natural language processing is the most conventional. Such a finding is consistent with what we show in the mixed-effect regression model. On the other hand, in Fig. S5. A, we show that machine learning is the most citation-disruptive field, while data mining and natural language processing are less disruptive.

**Qualitative analysis of top-cited, citation-disruptive, and novel papers**

In this section, we ranked the top papers by different metrics we used for analysis (bottom papers in the case of atypicality and conventionality), including citation count within five years of publication (C5), citation disruptive index (dc), atypicality, and conventionality. Then, we rank publications by combining the scaled variables with a min-max standardization we used in the previous lists. For each of the variables, we scaled them with a min-max scaler, keeping them between 0 and 1 before combining them to create a standardized index. We subtract the scaled atypicality and conventionality to get the standardized value since the smaller for these two measures, the better. We first show the ranking with a standardized index with scaled C5, dc, atypicality, and conventionality. Then, we show the ranking with scaled C5, atypicality, and conventionality.

Color code:

**Teams with all academic authors**

**Teams with all industry authors**

**Academic and industry collaboration teams**

Bottom 10 Atypical (e.g., highly novel work):

1. Learn To Be Uncertain Leveraging Uncertain Labels In Chest X Rays With Bayesian Neural Networks
2. Hydrological Time Series Prediction Model Based On Attention Lstm Neural Network
3. Software Malpractice In The Age Of AI A Guide For The Wary Tech Company
4. Symmetry Detection And Exploitation For Function Approximation In Deep Rl
5. Value Aware Loss Function For Model Based Reinforcement Learning
6. Understanding Measures Of Uncertainty For Adversarial Example Detection
7. Model Asset Exchange Path To Ubiquitous Deep Learning Deployment
8. Inception V3 Based Recommender System For Crops
9. Walking With Mind Mental Imagery Enhanced Embodied QA
10. Application Of Structured Illumination In Nano Scale Vision

Bottom 10 Conventional (e.g., highly novel work):
1. Deep Learning For Climate Models Of The Atlantic Ocean
2. Training Autoencoders In Sparse Domain
3. Generalized Representation Learning Methods For Deep Reinforcement Learning
4. Cause Learning Granger Causality From Event Sequences Using Attribution Methods
5. An Empirical Evaluation Of Annotation Practices In Corpora From Language Documentation
6. Mer Dimes A Planetary Landing Application Of Computer Vision
7. Intelligent Detection And Recognition System For Mask Wearing Based On Improved Retinaface Algorithm
8. Action Conditioned Convolutional Future Regression Models For Robot Imitation Learning
9. The Guiding Role of Reward Based on Phased Goal in Reinforcement Learning
10. GNNExplainer: Generating Explanations for Graph Neural Networks

Top 10 Citation-Disruptive (e.g., highly impactful work)

1. Adam A Method For Stochastic Optimization
2. Very Deep Convolutional Networks For Large Scale Image Recognition
3. Attention Is All You Need
4. The Pagerank Citation Ranking Bringing Order To The Web
5. Pytorch An Imperative Style High Performance Deep Learning Library
6. Explaining And Harnessing Adversarial Examples
7. Efficient Estimation Of Word Representations In Vector Space
8. A Comparative Study On Feature Selection In Text Categorization
9. Deep Compression Compressing Deep Neural Networks With Pruning Trained Quantization And Huffman Coding
10. Multi-Scale Context Aggregation by Dilated Convolutions

Top standardized index ($C5_{min-max}$ + $dc_{min-max}$ - $atypicality_{min-max}$ - $conventionality_{min-max}$) (e.g., highly impactful and novel work):

1. U Net Convolutional Networks For Biomedical Image Segmentation
2. Ros An Open Source Robot Operating System
3. Going Deeper With Convolutions
4. Frame Compatible Formats For 3d Video Distribution
5. Fully Convolutional Networks For Semantic Segmentation
6. Freesound Technical Demo
7. Annotea An Open Rdf Infrastructure For Shared Web Annotations
8. Building A Test Collection For Complex Document Information Processing
9. Imagenet Classification With Deep Convolutional Neural Networks
10. Neural Machine Translation By Jointly Learning To Align And Translate

Top standardized index ($C5_{min-max}$ - $atypicality_{min-max}$ - $conventionality_{min-max}$ (e.g., highly impactful and novel work):

1. Going Deeper With Convolutions
2. U Net Convolutional Networks For Biomedical Image Segmentation
3. Fully Convolutional Networks For Semantic Segmentation
4. Neural Machine Translation By Jointly Learning To Align And Translate
5. Glove Global Vectors For Word Representation
6. Imagenet Classification With Deep Convolutional Neural Networks
7. Generative Adversarial Nets
8. Faster R Cnn Towards Real Time Object Detection With Region Proposal Networks
9. Rich Feature Hierarchies For Accurate Object Detection And Semantic Segmentation
10. Caffe Convolutional Architecture For Fast Feature Embedding